\documentclass[aps,twocolumn,nofootinbib]{revtex4}
\usepackage{amssymb}
\usepackage{amsmath}
\usepackage{graphicx}
\usepackage[sort&compress]{natbib}

\bibpunct[, ]{}{}{,}{s}{,}{,}

\begin{document}

\title{Sliding grafted polymer layers}
\author{Vladimir A. Baulin, Albert Johner$^{\dag }$ and Carlos M. Marques$^{\dag }$}
\email{marques@ics.u-strasbg.fr}
\affiliation{Institut Charles Sadron, 67083
Strasbourg Cedex, France \vskip 0.1cm and $^\dag $Laboratoire Europ\'{e}en
Associ\'{e}, ICS (Strasbourg)/Max-Planck-Institute f\"{u}r
Polymerforschung (Mainz)}

\begin{abstract}
We study theoretically the structure of sliding grafted polymer layers or
SGP layers. These interfacial structures are built by attaching each polymer
to the substrate with a ring-like molecule such as cyclodextrins. Such a
topological grafting mode allows the chains to freely slide along the
attachment point. Escape from the sliding link is prevented by bulky capping
groups. We show that grafts in the mushroom regime adopt mainly symmetric
configurations (with comparable branch sizes) while grafts in dense layers
are highly dissymmetric so that only one branch \emph{per} graft
participates in the layer. Sliding layers on small colloids or star-like
sliding micelles exhibit an intermediate behavior where the number of longer
branches participating in the corona is independent of the total number of
branches. This regime also exists for sliding surface-micelles comprising
less chains but it is narrower.
\end{abstract}

\maketitle

\section{Introduction}

Rotaxanes are molecular complexes formed when a ring like molecule, the
rotor, is threaded over a linear molecule, the rotating axis.\cite{Nakashima}
The polymeric versions of rotaxanes are named polyrotaxanes. These necklace
structures are built by threading several ring molecules over a polymer
chain. Unthreading is also prevented by subsequent capping of the chain ends.%
\cite{Ogino,OginoOhata} Although the usual chemical and physical forces are
also at work in polyrotaxanes, the peculiar character of these complexes is
determined by the topological nature of each of its components. Such
materials are thus also known as topological materials. Polyrotaxanes are
being intensely scrutinized for advanced specific applications as molecular
shuttles, "insulated molecular wires", supramolecular light-harvesting
antenna systems or sliding gels.\cite%
{PGGSliding,Cacialli,Ballardini,HaradaAcc,Schalley,Tamura} They can be made
from different linear polymers\cite{HaradaCoo,Wei} combined with different
cyclic molecules, in different solvents.\cite{Rekharsky} One of the most
well studied systems involve poly(ethylene-oxide) and cyclodextrins, which
are oligosaccharides of $6$, $7$ or $8$ glucose units assembled as rings.
Although in most cases polyrotaxanes are formed with a very high density of
cyclodextrins threaded over the polymer chain, recent strategies for complex
formation\cite{Shridhar} allow for only one or a low number of cyclodextrins
per chain. The cyclodextrin can then further be grafted to a surface,
resulting in a grafted polymer layer where the chains retain the ability to
slide through the grafting ring. We coin the acronym SGP layers for such
structures, standing for sliding grafted polymer layers.

Layers of grafted polymers have a wide range of applications,\cite{CohenSt}
ranging from the colloidal stabilization of industry formulations, water
treatment and mineral recovery, to the control of surface wetting and
adhesion or to the protection of stealth liposomes from the human immune
system in drug delivery.\cite{Lipowsky} Sparsely grafted polymers are often
referred to as mushrooms, while more dense systems above the surface
overlapping density are known as brushes. Polymer theories for mushrooms and
brushes have been developed during the last two decades,\cite%
{SA,PGG,HTL,Milner} and their predictions successfully compared with elegant
experiments.\cite{Auvray} In this paper we revisit grafted polymer theories
introducing a key modification that will allow the polymer to be attached to
the surface in a sliding manner. As we shall see, this induces important
differences in the equilibrium and dynamic behavior of the layers, both in
the mushroom and brush regimes. In section \ref{secsingle} we consider ideal
sliding mushrooms of chains grafted with one or several sliding links.
Section \ref{secbrush} discusses denser layers and the crossover from
sliding mushrooms to sliding brushes is discussed in section \ref%
{sectransition}. In section \ref{secstars} we account for excluded volume
correlations and focus on sliding bulk and surface aggregates that embody
the sliding mushroom as a special case. The final section reviews our key
results and discuss their experimental relevance.

\section{SGP layers: Mushroom regime\label{secsingle}}

We study here SGP layers composed of isolated chains grafted to a planar
surface by a sliding link. We assume Gaussian statistics and will discuss
later in section \ref{secstars} excluded volume effects.

\subsection{Fixed sliding links}

\begin{figure}[]
\begin{center}
\includegraphics[width=2.5cm]{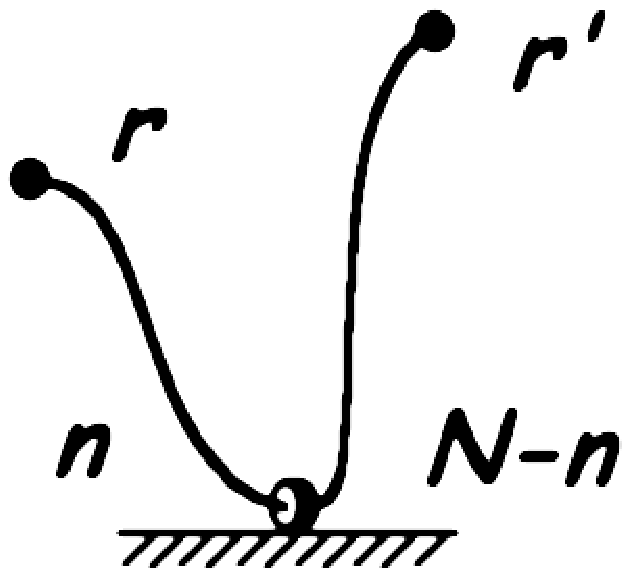} \vskip 1cm %
\includegraphics[width=5cm]{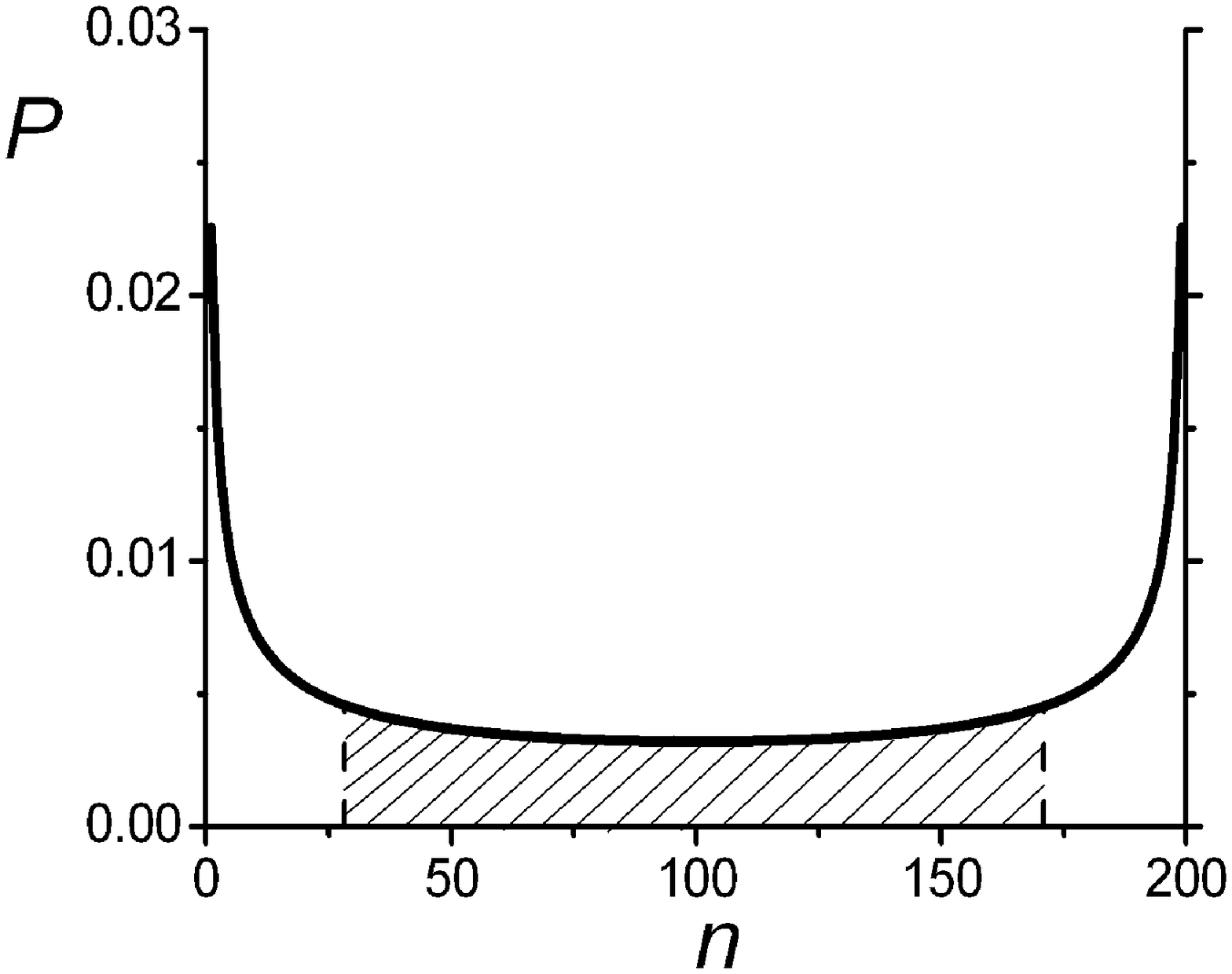}
\end{center}
\caption{Schematic picture and the lengths distribution function of a
Gaussian chain grafted to a surface by a sliding link. Two branches can
exchange monomers, while the total chain length is fixed $N=200$. The
central region which corresponds to symmetric configurations is defined by
the condition $\protect\int_{n}^{N-n}P(n)dn=1/2$ (hatched).}
\label{single}
\end{figure}

The sliding link is attached to a fixed position on the surface and allows
free exchange of monomers between the two branches of the chain with the
total number of monomers $N$, see Figure \ref{single}. Thus, one branch has $%
n$ monomers, the other branch has $N-n$ monomers. The Gaussian nature of the
two branches results in the absence of branch correlations and the Green
function\cite{Doi} of a chain is the product of the Green functions of the
two branches. Let the grafting point be at $\mathbf{a}=\{0,0,a\}$, where $a$
is a monomer size. The total Green function reads

\begin{equation}
G(\mathbf{r},\mathbf{r}^{\prime })=G_{n}(\mathbf{r},\mathbf{a})G_{N-n}(%
\mathbf{a},\mathbf{r}^{\prime }),  \label{G1}
\end{equation}%
where $\mathbf{r}$ and $\mathbf{r}^{\prime }$ are the coordinates of free
ends. The Green functions factorize over the directions $x$, $y$ and $z$,
\textit{i.e.} $G_{n}(\mathbf{r},\mathbf{r}^{\prime })=G_{n}^{x}(x,x^{\prime
})G_{n}^{y}(y,y^{\prime })G_{n}^{z}(z,z^{\prime })$. In the $x$ and $y$
directions the Green functions retain then bulk structure: $%
G_{n}^{x}(x,x^{\prime })=\left( 3/(2\pi na^{2})\right) ^{1/2}\exp \left[ -%
\frac{3}{2na^{2}}(x-x^{\prime })^{2}\right] $ and for a similar term for $%
G_{n}^{z}(z,z^{\prime })$ within the $z$ direction one needs to account for
the impermeability of the wall:

\begin{eqnarray}
G_{n}^{z}(z,z^{\prime })&=&\left( \frac{3}{2\pi na^{2}}\right) ^{1/2}\left[
\exp (-\frac{3}{2na^{2}}(z-z^{\prime })^{2})\right.  \notag \\
&&\left.-\exp (-\frac{3}{2na^{2}}(z+z^{\prime })^{2})\right]  \label{gz}
\end{eqnarray}

We focus first on the probability distribution function $P(n)$ describing
the number of configurations with branches of length $n$ and $N-n$. The
probability distribution $P(n)$ is calculated from the partition function $%
Z(n)=\int G(\mathbf{r},\mathbf{r}^{\prime })d\mathbf{r}d\mathbf{r}^{\prime }$%
, as $P(n)=Z(n)/Z$, with $Z=\int_{0}^{N}Z(n)dn$. In the limit where the
radius of gyration of each branch is larger then a monomer size, \textit{i.e.%
} $R_{g}(n)\sim a\sqrt{n}\gg a$, $P(n)$ can be written as

\begin{equation}
P(n)=\frac{1}{\pi \sqrt{n(N-n)}}  \label{idealsingle}
\end{equation}%
The equivalence of the two branches is reflected in the symmetry of this
function which have a minimum at $n=N/2$. This partition function is
dominated by \emph{symmetric} configurations, in the sense that it presents
only a weak divergence at $n=0$ which does not dominate its integral. Half
of the branches (Figure \ref{single}) belong to the central region $1/2-%
\sqrt{2}/4<n/N<1/2+\sqrt{2}/4$.

\begin{figure}[tbp]
\begin{center}
\includegraphics[width=7cm]{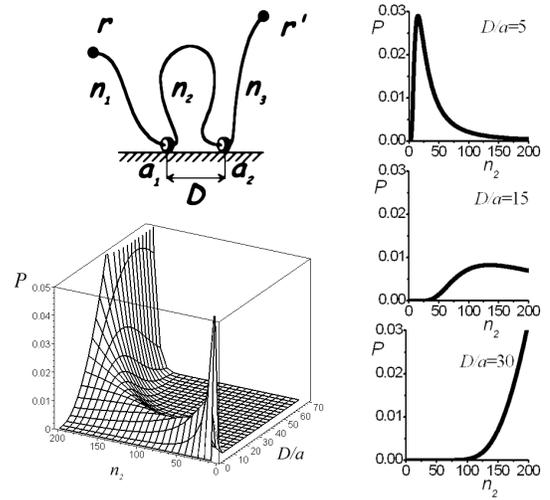}
\end{center}
\caption{Polymer chain grafted to a surface by two sliding links $\mathbf{a}%
_{1}$ and $\mathbf{a}_{2}$ separated by the distance $D$ between them and
the distribution of monomers in the loop $P(n_{2})$ for the total chain
length $N=200$. The 2D cuts for $D/a=5$, $15$ and $30$ are shown in the
inset.}
\label{double}
\end{figure}

We consider now a Gaussian chain grafted to a plane surface by \emph{two}
fixed sliding links at a distance $D$ between them. The position of two
grafting points are $\mathbf{a}_{1}=\{0,0,a\}$ and $\mathbf{a}_{2}=\{D,0,a\}$%
. The chain has two free ends comprising $n_{1}$ and $n_{3}$ monomers and
one middle loop with $n_{2}$ monomers, while the total number of monomers in
the chain is $N=n_{1}+n_{2}+n_{3}$. The Green function of the chain is

\begin{equation}
G(\mathbf{r},\mathbf{r}^{\prime })=G_{n_{1}}(\mathbf{r},\mathbf{a}%
_{1})G_{n_{2}}(\mathbf{a}_{1},\mathbf{a}_{2})G_{n_{3}}(\mathbf{a}_{2},%
\mathbf{r}^{\prime })  \label{G2}
\end{equation}

The integration over positions of the free ends, $\mathbf{r}$ and $\mathbf{r}%
^{\prime }$, gives as before the partition function for the tails and now
also for the loop. It can be expressed for instance as a function of the
number of monomers in the first tail, $n_{1}$, and the number of monomers in
the loop, $n_{2}$:

\begin{equation}
Z(n_{1},n_{2})\sim \frac{1}{\sqrt{n_{1}(N-n_{1}-n_{2})}}\frac{\exp \left[ -%
\frac{3D^{2}}{2n_{2}}\right] }{n_{2}^{3/2}}\left( 1-e^{-\frac{6}{n_{2}}%
}\right)
\end{equation}%
The first term expresses the usual free end contribution of the form $%
n^{-1/2}$, while the last two terms account for the loop. Since the two free
ends are identical, we concentrate on the probability distribution for the
monomers in the loop $P(n_{2})=Z(n_{2})/Z$, where the loop partition
function $Z(n_{2})=\int_{0}^{N-n_{2}}Z(n_{1},n_{2})dn_{1}$ is given by

\begin{figure}[b]
\begin{center}
\includegraphics[width=7cm]{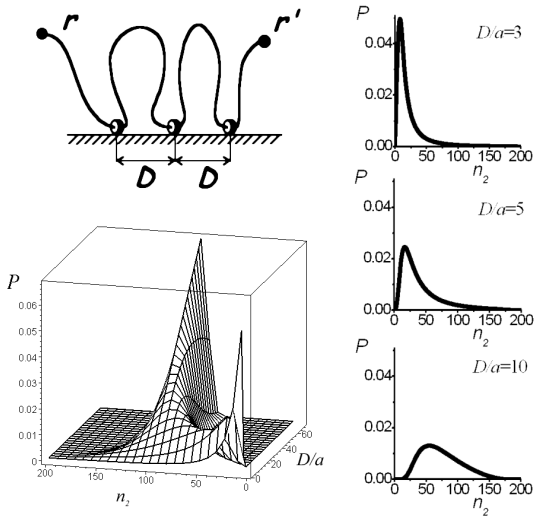}
\end{center}
\caption{Polymer chain grafted by three sliding links with a distance $D$
between them and the distribution of monomers in one of the loops $P(n_{2})$
for the total chain length $N=200$. The 2D cuts for $D/a=3$, $5$ and $10$
are shown in the inset.}
\label{triple}
\end{figure}
\begin{figure}[]
\begin{center}
\includegraphics[width=8cm]{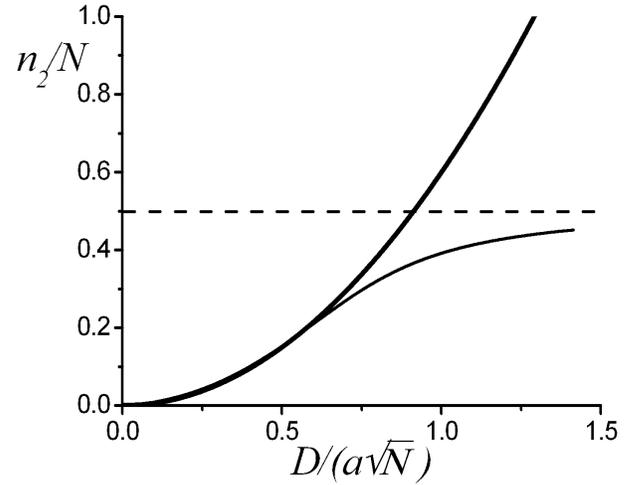}
\end{center}
\caption{Fraction of monomers in the loop $n_{2}/N$ corresponding to the
maximum of the distribution $P(n_{2})$ as a function of a scaled distance
between the grafting points $D/(a\protect\sqrt{N})$ for a chain grafted by
\textit{two} (thick) and \textit{three} (thin) sliding links. Compare with
Figures \protect\ref{double} and \protect\ref{triple}.}
\label{max}
\end{figure}


\begin{equation}
Z(n_{2})=\pi \frac{\exp \left[ -\frac{3D^{2}}{2n_{2}}\right] }{n_{2}^{3/2}}%
\left( 1-e^{-\frac{6}{n_{2}}}\right)
\end{equation}%
and the total partition function is $Z=\int_{0}^{N}Z(n_{2})dn_{2}$. A
three-dimensional plot of $P(n_{2})$ is presented in Figure \ref{double}. If
the distance $D$ between the links is large, most of the monomers are in the
loop and the sliding chain behaves as a chain fixed by two ends. For small
distances a loop is entropically unfavorable and the monomers are
distributed between the two ends. If the chain is grafted by three grafting
points (Figure \ref{triple}), the two loops turn out to be identical and the
monomers are distributed equally between them. Thus, for a single loop the
number of monomers in the loop corresponding to the maximum of $P(n_{2})$
goes to $N$ for large $D$, while for two loops it tends to $N/2$. This
implies the emergence of two loops of equal size (Figure \ref{max}).

In the general case of a chain grafted by $m$ sliding links separated by the
distances $D_{1,}D_{2},\ldots ,D_{m-1}$ along the $x$ direction such that $%
n_{k}\gg 1$, the partition function is the product

\begin{equation}
Z(n_{1},n_{2}\ldots n_{m+1})\sim \frac{1}{\sqrt{n_{1}n_{m+1}}}\prod_{k=2}^{m}%
\frac{\exp \left( -\frac{3\left( D_{k-1}/a\right) ^{2}}{2n_{k}}\right) }{%
n_{k}^{5/2}}  \label{Zn1}
\end{equation}%
which is completed by the condition of conservation of monomers $%
N=\sum_{k=1}^{m+1}n_{k}$. Since the total partition function is the
convolution integral over all variables, it can be calculated by Laplace
transform.\cite{foot1} This allows to calculate the long chain limit of the
total partition function

\begin{equation}
Z_{N\rightarrow \infty }\sim \prod_{k=2}^{m}\frac{1}{\left( D_{k-1}/a\right)
^{3}}  \label{Z}
\end{equation}

The structure of eq. (\ref{Zn1}) suggests that different loops are
equivalent to each other. In particular, if the distances between grafting
points are equal, $D_{k-1}=D$, the monomers should be equally distributed
between the loops.

\subsection{Sliding links with lateral mobility}

Let us turn to the situation where not only the chain can slide through the
grafting points, but the grafting points themselves can freely move on the
surface. This can be the case when cyclodextrins are grafted to the surface
of a liquid membrane. If a chain is grafted by a single mobile link, the
redistribution of monomers between two branches is the same as in the case
of a fixed grafting point (Figure \ref{single}). However, in case of several
mobile grafting points the distribution of monomers between free ends and
loops is changed.

\begin{figure}[]
\begin{center}
\includegraphics[width=6cm]{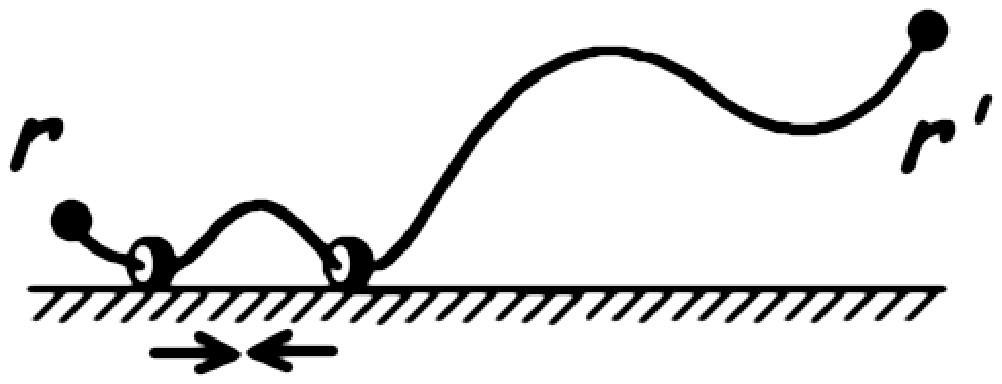} \vskip 1cm %
\includegraphics[width=7cm]{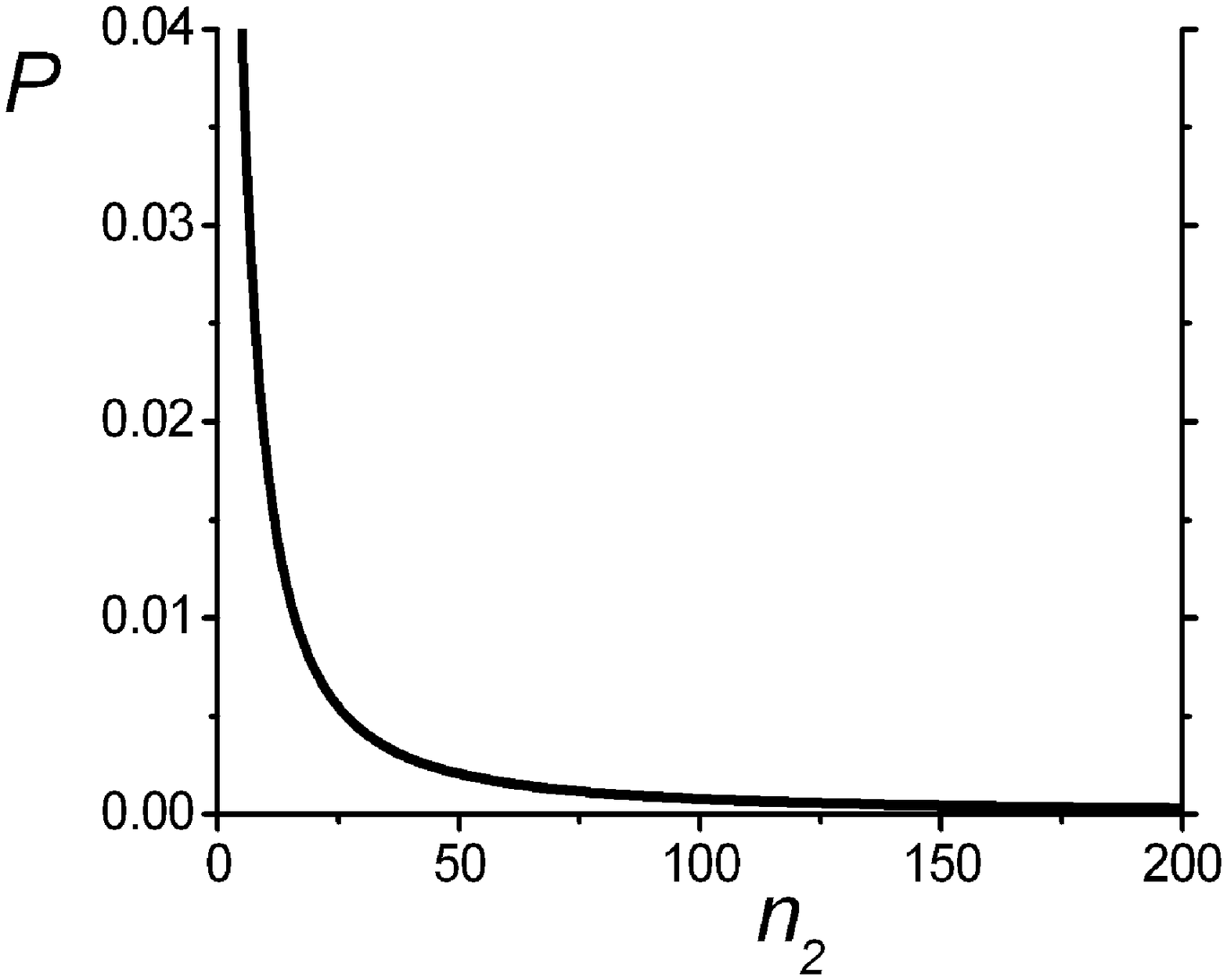}
\end{center}
\caption{Chain grafted by freely moving sliding links and corresponding
distribution of monomers in a loop $P(n_{2})$ for $N=200$.}
\label{sliding}
\end{figure}

Consider \textit{two} mobile sliding links placed at $\mathbf{a}%
_{1}=\{x_{1},y_{1},a\}$ and $\mathbf{a}_{2}=\{x_{2},y_{2},a\}$. We can use
again the expression of the Green function for fixed grafting points (\ref%
{G2}) but now the partition function is obtained by the integration both on
positions of free ends and positions of sliding links: $Z(n_{1},n_{2})=\int
G($\textbf{$r$}$,$\textbf{$r$}$^{\prime })d\mathbf{r}d\mathbf{r}^{\prime }d%
\mathbf{a}_{1}d\mathbf{a}_{2}$

\begin{equation}
Z(n_{1},n_{2})=\frac{1}{\sqrt{n_{1}n_{2}(N-n_{1}-n_{2})}}\left( 1-e^{-\frac{6%
}{n_{2}}}\right)
\end{equation}

Integration over the tails $n_{1}$ gives the distribution function of
monomers in the loop

\begin{equation}
P(n_{2})=\frac{1}{Z}\frac{\pi }{n_{2}^{3/2}},\text{ \ \ \ \ \ \ \ \ }n_{2}>1
\end{equation}%
where $Z=\int_{0}^{N}Z(n_{2})dn_{2}$ is the total partition function. $%
P(n_{2})$ is presented in Figure \ref{sliding}. It rapidly decreases with
the increasing size of the loop $n_{2}$, which shows that large loops are
not favorable.

In general, when a sliding chain is grafted by several mobile sliding links
the system acquires two additional degrees of freedom (two transversal
coordinates). Each of them contributes to the partition function as $\sqrt{%
n_{k}}$. Thus,

\begin{equation}
Z(n_{1},n_{2},\ldots ,n_{m+1})\sim \frac{1}{\sqrt{n_{1}n_{m+1}}}%
\prod_{k=2}^{m}\frac{1}{n_{k}^{3/2}}
\end{equation}%
Applying the Laplace transform to the total partition function we get $Z\sim
(1/\sqrt{N})^{(m-1)}$. The form of the partition function brings us to the
same conclusion: the system prefers to eliminate the loops. This leads to an
effective entropic attraction between mobile grafts, which tend to stick
together even in the absence of any additional forces.

\begin{figure}[b]
\begin{center}
\includegraphics[width=8cm]{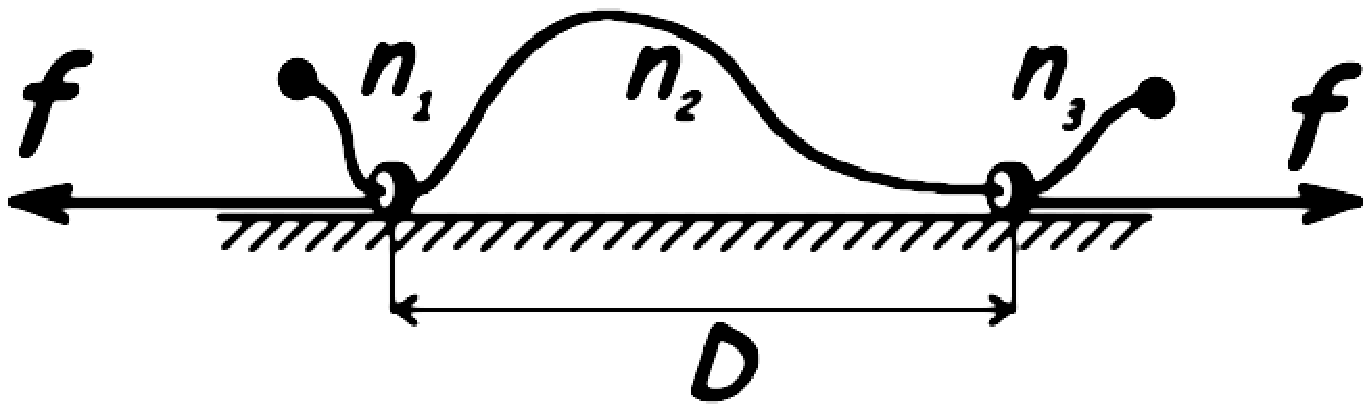} \vskip 1cm %
\includegraphics[width=6.5cm]{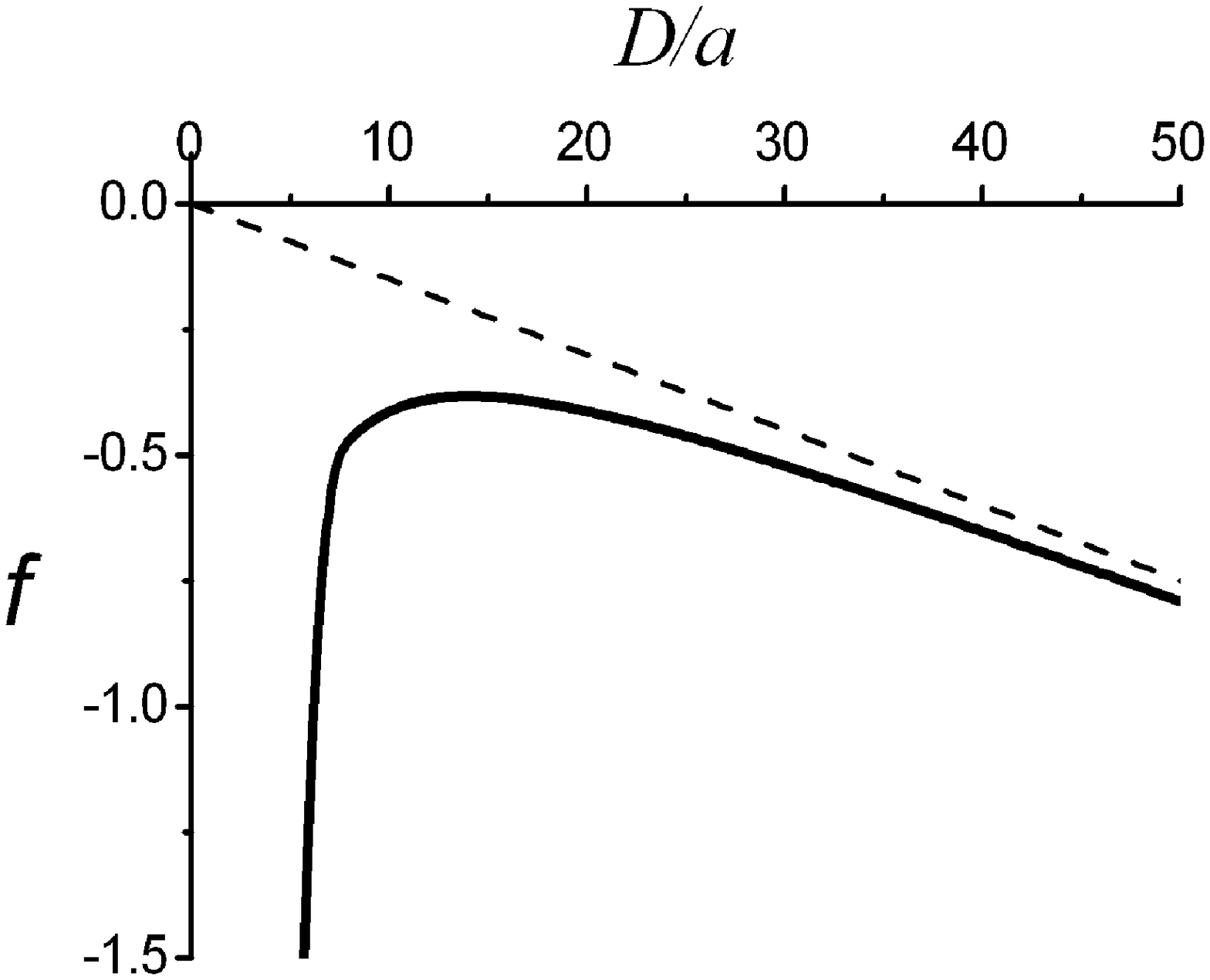}
\end{center}
\caption{Extension force $\mathbf{f}$ of a chain grafted by two sliding
links as a function of a distance $D$ between them (solid) in comparison to
the extension of the end-grafted Gaussian chain of the same length (dash).
The total chain length is $N=200$.}
\label{ext}
\end{figure}

\subsection{Sliding grafted chain under a pulling force}

Once the sliding links are stuck together, one need to apply the force to
separate them. In figure \ref{ext} we plot the force in units of $kT$ needed
to separate the two links to the distance $D$: $f=\partial \ln Z/\partial D$%
. At large distances, $D^{2}/(Na^{2})\gg 1$, the force coincides with that
of the chain grafted by two ends, $f=-3D/(Na)$. At small distances, $%
D^{2}/(Na^{2})\ll 1$, the curve has a logarithmic divergence, $f\sim \ln
(D/a)$. The curve passes by a maximum coinciding with the creation of the
loop.

We can see the difference between a chain grafted by one end to a surface
and a chain grafted by a sliding link also when one applies a force parallel
to the surface to one of free ends $\mathbf{f}=\{f_{x},0,0\}$. The partition
function of a sliding grafted chain under a pulling force is

\begin{equation}
Z(\mathbf{r}^{\prime })=\int_{0}^{N}dn\int d\mathbf{r}G_{n}(\mathbf{r},%
\mathbf{a})G_{N-n}(\mathbf{a},\mathbf{r}^{\prime })e^{f_{x}r_{x}^{\prime }}
\end{equation}%
and the total partition function $Z=\int Z(\mathbf{r}^{\prime })d\mathbf{r}%
^{\prime }$is

\begin{equation}
Z=6e^{\beta /2}I_{0}\left( \beta /2\right) ,
\end{equation}%
where $\beta =\frac{Na^{2}}{6}f_{x}^{2}$ is a scaling parameter associated
with the magnitude of the applied force $\mathbf{f}$ and $I_{0}(x)$ is the
zero-order Bessel I-function.\cite{Abramowitz} At the same time, the
corresponding partition functions of an end-grafted chain, denoted by the
subscript $0$, are $Z_{0}(\mathbf{r}^{\prime })=G_{N}(\mathbf{a},\mathbf{r}%
^{\prime })e^{f_{x}r_{x}^{\prime }}$and $Z_{0}\sim e^{\beta }/\sqrt{N}$ for
large $N$.

\begin{figure}[]
\begin{center}
\includegraphics[width=7cm]{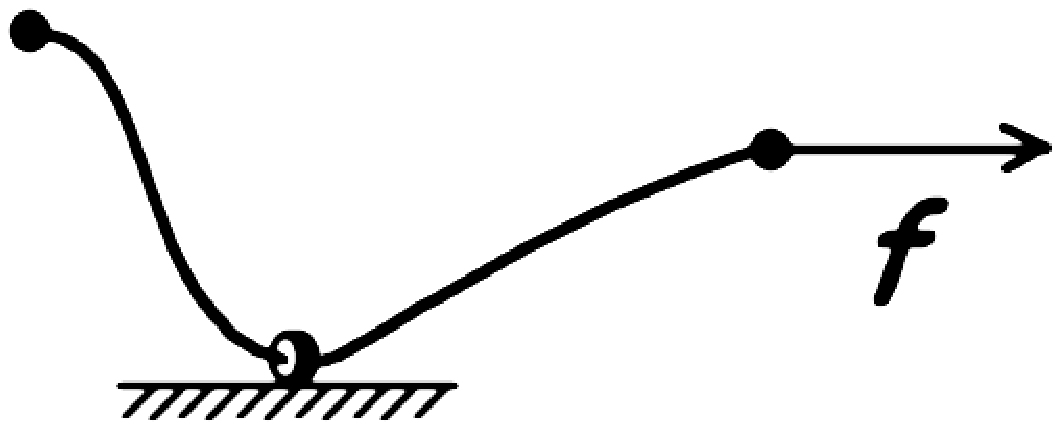} \vskip 0.8cm %
\includegraphics[width=7cm]{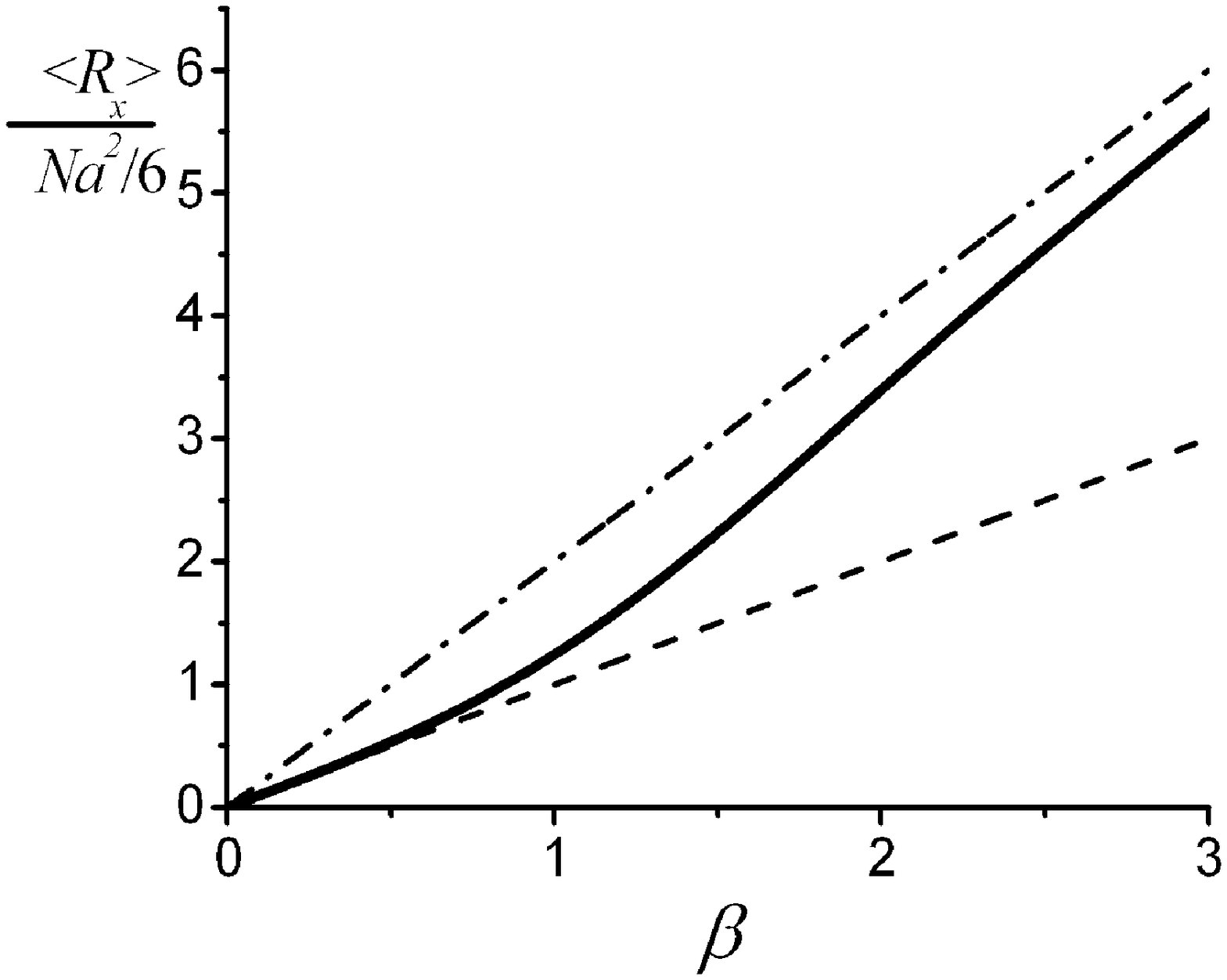}
\end{center}
\caption{Transversal dimension of a \textit{sliding chain} $\frac{%
\left\langle R_{x}\right\rangle }{\protect\sqrt{Na^{2}/6}}$ as a function of
the scaled force $\protect\beta =f_{x}\protect\sqrt{Na^{2}/6}$ applied to a
free end (solid) in comparison with the dimension of the \textit{end-grafted
chain} of the same length $N$ (dash dot) and two times shorter chain $N/2$
(dash).}
\label{f1}
\end{figure}

For a relatively large force $f_{x}$ both end-grafted and sliding chains are
fully stretched with comparable configurations. However, for relatively
small $f_{x}$ a sliding chain prefers symmetric configurations with two
branches of more or less equal size. In this limit the sliding chain
resembles two end-grafted chains comprised of $N/2$ monomers. To illustrate
such behavior we compare the average distance from the grafting point in the
$x$-direction for both chains. The average distance for an end-grafted chain
is $\left\langle R_{x}\right\rangle _{0}=\frac{Na^{2}}{3}f_{x}$, while the
average distance for a sliding chain is

\begin{equation}
\left\langle R_{x}\right\rangle =\frac{\left\langle R_{x}\right\rangle _{0}}{%
2}\left( 1+\frac{I_{1}(\frac{\beta }{2})}{I_{0}(\frac{\beta }{2})}\right)
\end{equation}%
The resulting curves are plotted in Figure \ref{f1}. The increasing force
provokes the transition of a sliding chain from a symmetric configuration
with two branches of size $N/2$ to fully stretched with a single branch of
length $N$.

The same behavior is expected for the average square distance from the
grafting point. In the case of an end-grafted chain we obtain $\left\langle
R_{x}^{2}\right\rangle _{0}=\frac{Na^{2}}{3}\left( 1+2\beta \right) $, while
for a sliding chain

\begin{equation}
\left\langle R_{x}^{2}\right\rangle =\frac{\left\langle
R_{x}^{2}\right\rangle _{0}}{2}\left( 1-\frac{1-2\beta }{1+2\beta }\frac{%
I_{1}(\frac{\beta }{2})}{I_{0}(\frac{\beta }{2})}\right)
\end{equation}

\begin{figure}[b]
\begin{center}
\includegraphics[width=7cm]{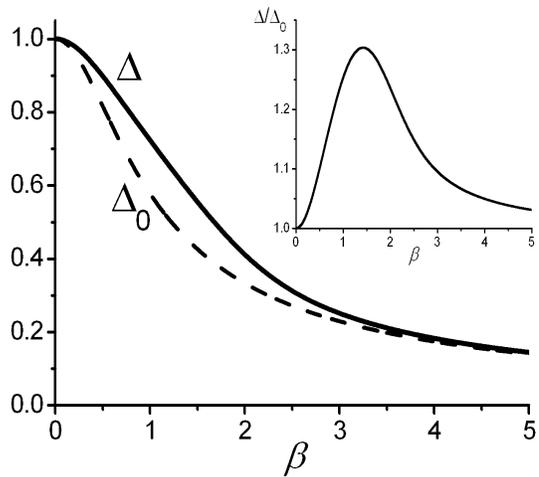}
\end{center}
\caption{The dispersion $\Delta =\protect\sqrt{(\left\langle
R_{x}^{2}\right\rangle -\left\langle R_{x}\right\rangle ^{2})/\left\langle
R_{x}^{2}\right\rangle }$ of a sliding chain (solid) and an end-grafted
chain $\Delta _{0}$ of the same length (dash) under a pulling force applied
to a free end. Inset: the relation $\Delta /\Delta _{0}$ between dispersions
of sliding and end-grafted chains.}
\label{fluct}
\end{figure}

A sliding chain has more degrees of freedom as compared to an end-tethered
chain. The \textit{dispersion} of the size

\begin{equation}
\Delta (\beta )=\sqrt{\frac{\left\langle R_{x}^{2}\right\rangle
-\left\langle R_{x}\right\rangle ^{2}}{\left\langle R_{x}^{2}\right\rangle }}
\end{equation}%
of the chain under a pulling force is larger than that of the end-grafted
chain of the same length (Figure \ref{fluct}). As expected, the dispersion
of a sliding chain coincide with the dispersion of an end-grafted chain when
the chains are very stretched.

\section{SGP layers: The brush regime\label{secbrush}}

\begin{figure}[]
\begin{center}
\includegraphics[width=7cm]{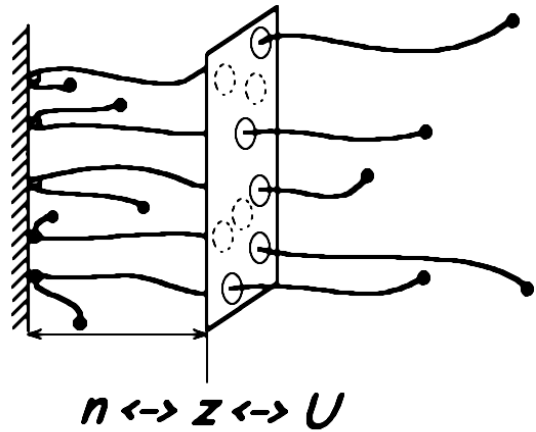}
\end{center}
\caption{Schematic picture of a brush of sliding polymers.}
\label{brush}
\end{figure}

When the grafting density in the SGP layers is high enough the different
chains and chain branches will interact strongly. Each chain can exchange
monomers between two branches. Although lengths of individual chains are
equal, lengths of their branches can vary from chain to chain. Hence, one
can assume that the branches of the sliding chains are independent "chains"
of annealed length composing a brush with annealed polydispersity (Figure %
\ref{brush}). We will treat a polydisperse brush in the framework of the
self-consistent field theory of brushes in the strong stretching regime\cite%
{Milner} valid for high grafting densities. In this limit the configurations
of chains are considered as trajectories $z(n)$ of effective "particles"
moving in the field $U$ which depends only on the distance from the grafting
surface. Thus, the molecular weight of a chain $n$ is analogous to the time
needed for a "particle" for traveling from any distance $z_{0}$ to the
grafting surface $z=0$. This analogy leads to a Newton equation of motion
for $z(n)$, the distance from the grafting surface
\begin{equation}
\frac{d^{2}z}{dn^{2}}=\frac{dU}{dz}.  \label{eqmot}
\end{equation}%
This allows to relate the distance $z$, the molecular weight of a chain $n$
and the self-consistent potential $U$. We consider all three variables to be
dimensionless.

Let $\sigma (n)$ be the number of chains per unit area with molecular weight
$n$. Then $S(n)=\int_{n}^{N}\sigma (n)dn$ is the number per unit area of
chains with molecular weight larger than $n$. The total number of chains per
unit area is $S_{0}\equiv S(n=0)$. The polydispersity $\sigma (n)$ is
related to the distribution of chain lengths $P(n)$. Thus, we can write $%
S(n)=S_{0}\int_{n}^{N}P(n)dn$, where $P(n)$ is normalized: $%
\int_{0}^{N}P(n)dn=1$. This expression can be rewritten as

\begin{equation}
S(n)=S_{0}\int_{z}^{H}P(z)dz=S_{0}\int_{0}^{U}P(U^{\prime })dU^{\prime }
\label{Sn}
\end{equation}

The concentration at a given height $z$ (or potential $U$) is constructed by
chains whose ends start at larger heights $z^{\prime }$ (lower potentials $%
U^{\prime }<U$):
\begin{eqnarray}
\phi (z(U)) &=&S_{0}\int_{0}^{U}P(U^{\prime })\frac{dn}{dz}(U,U^{\prime
})dU^{\prime }=  \notag \\
&&S_{0}\int_{0}^{U}\frac{P(U^{\prime })dU^{\prime }}{\sqrt{2(U-U^{\prime })}}
\label{phi}
\end{eqnarray}%
Here the expression for $dz/dn$ is obtained from the integration of eq. (\ref%
{eqmot}).

We will write the potential $U$ in the form $U=\frac{w^{2}}{2}\phi ^{2}$,
which corresponds to a $\theta $-solution, where the mean field
approximation is justified. Here $w^{2}$ is the effective third virial
coefficient. This form will allow us to get analytical results, which remain
qualitatively correct also for a more conventional choice assuming two body
interactions (good solvent). We can eliminate the volume fraction $\phi $
and write (\ref{phi}) as a closed equation for $P(U^{\prime })$:

\begin{equation}
\frac{\sqrt{2}}{w}\sqrt{U}=S_{0}\int_{0}^{U}\frac{P(U^{\prime })dU^{\prime }%
}{\sqrt{2(U-U^{\prime })}}  \label{UP}
\end{equation}

The solution of this equation, obtained using the Laplace transform\cite%
{foot2} is

\begin{equation}
U(n)=wS(n)  \label{Un}
\end{equation}

The chemical potential of a chain is a sum of the chemical potentials of the
two branches $\mu _{chain}=\frac{\mu (n)+\mu (N-n)}{2}$, where $\mu
(n)=\int_{0}^{N}U(n^{\prime })dn^{\prime }$. We can write $\mu _{chain}(n)$
as a functional of $P(n)$ which we integrate by parts using the symmetry of $%
P(n)$

\begin{eqnarray}
\mu _{chain} &=&\frac{wS_{0}}{2}\left[ \int_{0}^{n}dn^{\prime
}\int_{n^{\prime }}^{N}P(n^{\prime \prime })dn^{\prime \prime }+\right.
\notag \\
&&\left.\int_{0}^{N-n}dn^{\prime }\int_{n^{\prime }}^{N}P(n^{\prime \prime
})dn^{\prime \prime }\right]  \notag \\
&=&\frac{wS_{0}}{2}\left[ \int_{0}^{n}\left( N-n+n^{\prime }\right)
P(n^{\prime })dn^{\prime }+\right.  \notag \\
&&\left.\int_{n}^{N}\left( N-n^{\prime }+n\right) P(n^{\prime })dn^{\prime }
\right]
\end{eqnarray}

Minimization of the free energy functional

\begin{equation}
F\left\{ P(n)\right\} ={\frac{S_{0}^{2}}{2}}\int_{0}^{N}P(n)\mu
_{chain}\left\{ P(n)\right\} dn
\end{equation}%
with respect to $P(n)$ along with the normalization condition $%
\int_{0}^{N}P(n)dn=1$ gives the equilibrium distribution of chain lengths:

\begin{equation}
P(n)=\left\{
\begin{array}{c}
\frac{1}{2}\delta (n),\text{ \ \ \ \ \ \ \ \ \ \ \ \ }0<n<\frac{N}{2} \\
\frac{1}{2}\delta (N-n),\text{ \ \ \ \ }\frac{N}{2}<n<N%
\end{array}%
\right.  \label{strongstretch}
\end{equation}%
In a densely grafted layer the chains adopt very dissymmetric configurations
and behave as end-grafted chains. The strong stretching approximation
ignores local density fluctuations in the layer. We expect the
delta-functions in eq.(\ref{strongstretch}) to stand for localized functions
decaying over one correlation length (\emph{blob} size). This we consider
next.

\section{Transition from sliding mushrooms to sliding brushes\label%
{sectransition}}

\begin{figure}[]
\begin{center}
\includegraphics[width=4cm]{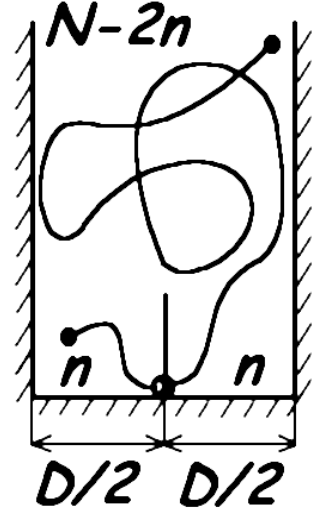}\vskip 0.7cm %
\includegraphics[width=7cm]{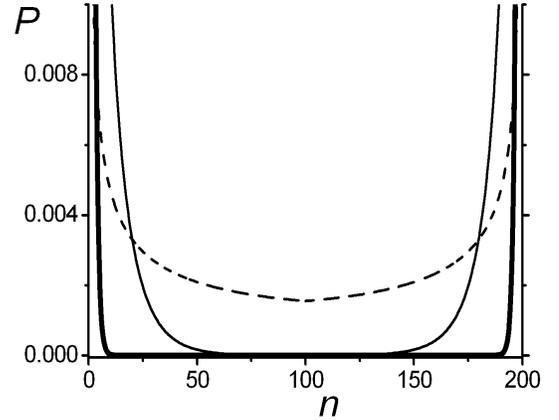}
\end{center}
\caption{Lengths distribution function of a sliding chain in a box for $N=200
$ and the wall-to-wall distance $D/a=80$ (dash), $D/a=15$ (thin) and $D/a=5$
(thick).}
\label{box}
\end{figure}

As we have seen, a single Gaussian chain grafted by a sliding link prefers
symmetric configurations, while sliding chains in a densely grafted brush
adopt stretched asymmetric configurations. Thus, there must be a crossover
region between the two configurations as the grafting density is increased.
Such an intermediate situation can be modeled as a Gaussian chain in a box.
The walls of the box mimic the steric repulsion of neighboring chains and
the decreasing distance between the walls models the increasing grafting
density. To model the steric repulsion between two branches of the same
chain we place a wall with a height equal to the size of the shortest branch
in the middle of the box as depicted in Figure \ref{box}. Assume that the
shortest branch has $n$ monomers and the distance between the walls is $D$.
The shortest branch and the part of the longest branch of length $n$ are
confined in smaller boxes of width $D/2$, while the rest of the longest
branch of length $N-2n$ is in the box of width $D$.

Such intermediate regime corresponds to $Na^{2}>D^{2}$ and $na^{2}<D^{2}$.
In this limits, the partition function of the chain confined in the box is

\begin{equation}
Z=Z_{\parallel }^{4}(n,\frac{D}{2})Z_{\parallel }^{2}(N-2n,D)Z_{\perp
}(n,a)Z_{\perp }(N-n,a)
\end{equation}%
where the perpendicular component is

\begin{equation}
Z_{\perp }(n,a)=\text{erf}\left( \sqrt{\frac{3}{2n}}\right) \sim \frac{1}{%
\sqrt{n}}
\end{equation}%
and the component parallel to the grafting surface is\cite{Doi}

\begin{equation}
Z_{\parallel }(n,\frac{D}{2})=\frac{4}{\pi }\sum_{p=1}^{\infty }\frac{1}{p}%
\sin ^{3}\left( \frac{\pi p}{2}\right) \exp \left( -\frac{\pi ^{2}p^{2}na^{2}%
}{6D^{2}}\right)
\end{equation}%
Since $na^{2}<D^{2}$ we can approximate this expression by the first mode

\begin{equation}
Z_{\parallel }(n,\frac{D}{2})\sim \exp \left( -\frac{\pi ^{2}na^{2}}{6D^{2}}%
\right)
\end{equation}

Thus, the partition function has the form

\begin{equation}
Z\sim \frac{\exp \left( -\frac{2\pi ^{2}na^{2}}{D^{2}}\right) }{\sqrt{%
n\left( N-n\right) }},
\end{equation}%
which is valid for $n$ corresponding to the short branch. Normalization of
this function leads to the following expression for the distribution of the
ends

\begin{eqnarray}
&&P(n)=\frac{1}{2\pi I_{0}\left( \pi ^{2}\frac{Na^{2}}{D^{2}}\right) }\times
\notag \\
&&\frac{1}{\sqrt{n(N-n)}}\left\{
\begin{array}{c}
\exp \left( 2\pi ^{2}\frac{(\frac{N}{2}-n)a^{2}}{D^{2}}\right) ,\text{ \ \ }%
0<n<\frac{N}{2} \\
\exp \left( 2\pi ^{2}\frac{(n-\frac{N}{2})a^{2}}{D^{2}}\right) ,\text{ \ \ }%
\frac{N}{2}<n<N%
\end{array}%
\right.
\end{eqnarray}

In the limit $Na^{2}\gg D^{2}$ and $na^{2}\ll D^{2}$ this function can be
approximated by

\begin{equation}
P(n)\sim \frac{a}{D}\frac{1}{\sqrt{n}}
\end{equation}%
This gives the estimate of the crossover value for the length the shorter
branch: $\sqrt{n^{\ast }}\sim D/a$. Above the overlap grafting density, the
size distribution of the branches is bimodal, the shorter branch comprising
of order $D^{2}/a^{2}$ monomers counts one \emph{blob}. When the overlap
density is approached from above, the size distribution spreads over the
whole interval and more symmetric configurations are favored (Figure \ref%
{box}).

\section{SGP layers in curved geometries: Stars and micelles\label{secstars}}

We consider now the case where the sliding links that anchor the polymer are
attached to curved surfaces, with radii much smaller than the unperturbed
chain size. This might be the case, for instance, if cyclodextrin rings are
attached on a packed configuration, resulting in a star-like polymer. More
commonly, this would also be the result of the micellization of amphiphilic
molecules carrying cyclodextrins as the headgroups. In any of these cases,
the resulting star-like object is composed of a fixed number of arms with
annealed lengths. Notice that such a bulk configuration can also arise at
interfaces if ring association takes place close to impenetrable wall. In
this chapter we first describe the partition function of the usual three
dimensional star and adapt such description for the annealed case, then
extend our results to the case of a surface star.

The partition function $Z_{p}$ of a star\cite{Duplantier} with $p$ equal
branches of contour length $N$ is given by the critical exponent $\gamma
_{p} $, $Z_{p}=N^{\gamma _{p}-1}$. Because a two arm star is also a linear
chain, one must have $\gamma _{1}=\gamma _{2}$.

Let us now consider a star with two arms of length $n_{1}$ and $n_{2}>n_{1}$%
, the partition function obeys the general scaling form $n_{1}^{\gamma
_{2}-1}(n_{2}/n_{1})^{x}$. In the limiting case $n_{1}\sim 1$, the one arm
partition function should be recovered, hence $x=\gamma _{1}-1$. This is now
generalized to an arbitrary star.\cite{Diez}

For a star with polydispersed arms all of different sizes, ranging from the
smallest $n_{1}$ to the largest $n_{p}$ the partition function can be
constructed step by step. Let first all $p$ arms have the size $n_{1}$ , the
partition function is $n_{1}^{\gamma _{p}-1}$, let now all chains but one
grow to the next size $n_{2}$, the partition function becomes $n_{1}^{\gamma
_{p}-1}n_{2}^{\gamma _{p-1}-1}/n_{1}^{\gamma _{p-1}-1}$, in the next step
let all outer chains but one grow to the next size $n_{3}$ and so on. As a
result:
\begin{equation}
Z_{p}=n_{1}^{\gamma _{p}-\gamma _{p-1}}n_{2}^{\gamma _{p-1}-\gamma
_{p-2}}\ldots n_{p-1}^{0}n_{p}^{\gamma _{1}-1}  \label{evstar}
\end{equation}

Consider now a sliding aggregate comprising $q$ chains. Let us characterize
each chain by the smallest of its two arms, the largest being its complement
to $N$, and let $n_{1}$ be the smallest of this $q$ arms, all by definition
smaller than $N/2$. The partition function of the sliding aggregate reads:
\begin{eqnarray}
Z(q)&=&\int_{cut}^{\frac{N}{2}}dn_{q}n_{q}^{\gamma _{q+1}-\gamma
_{q}}(N-n_{q})^{\gamma _{q}-\gamma _{q-1}}\ldots \times  \notag \\
&&\int_{cut}^{n_{2}}dn_{1}n_{1}^{\gamma _{2q}-\gamma
_{2q-1}}(N-n_{1})^{\gamma _{1}-1}  \label{slidingaggr}
\end{eqnarray}

The behavior of $Z(q)$ depends on whether the integrals are dominated by the
upper or the lower boundary, the lower boundary being a monomeric cut off
length . Let $p^{\star }$ be the value of the index such as $\gamma
_{p^{\ast }}-\gamma _{p^{\ast }-1}>-1$ and $\gamma _{p^{\ast }+1}-\gamma
_{p^{\ast }}\leq -1$.

(i) If there are few chains \emph{per} aggregate ($2q\leq p^{\star }$) all
integrals are dominated by the upper boundary and hence by symmetric
configurations.
\begin{equation}
Z(q)\propto N^{q}N^{\gamma _{2q}-1}  \label{smallaggr}
\end{equation}%
This corresponds to a $2q$-arm star, the extra factor stands for the choice
of monomers located at the core.

(ii) In the opposite limit of many chains \emph{per} aggregate ($q\geq
p^{\star }$), all integrals are dominated by the lower boundary and hence by
very dissymmetric chain configurations.
\begin{equation}
Z(q)\propto N^{\gamma _{q}-1}  \label{largeaggr}
\end{equation}%
This corresponds to a $q$-arm star.

(iii) In the intermediate regime ($q<p^{\star }<2q$), there are essentially $%
2q-p^{\star }$ dissymmetric chains and hence $p^{\star }-q$ symmetric ones.
The aggregate is thus equivalent to a $p^{\star }$-star with an additional
factor accounting for the freedom of symmetric configurations.
\begin{equation}
Z(q)\propto N^{p^{\star }-q}N^{\gamma _{p^{\star }}-1}  \label{intermaggr}
\end{equation}

\begin{figure}[]
\begin{center}
\includegraphics[width=8cm]{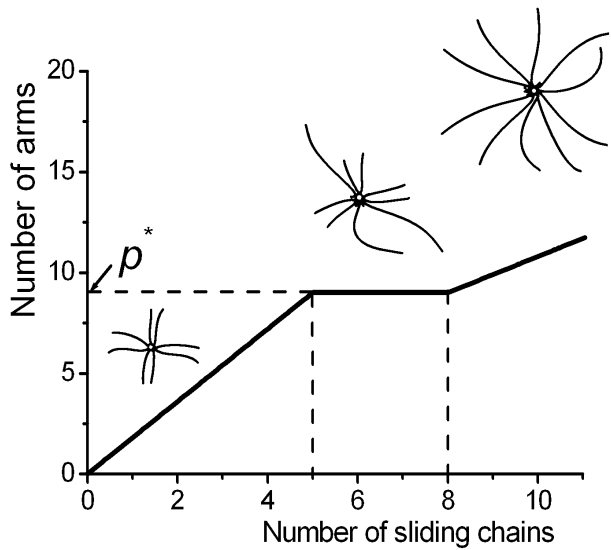}
\end{center}
\caption{Number of arms $p$ of a sliding bulk star as a function of the
number of sliding chains $q$. For a sliding surface star only the crossover
values are changed.}
\label{Narms}
\end{figure}

Star exponents $\gamma _{p}$ are known exactly in two dimensions and for
ideal chains ($d>4$). Otherwise first order $\epsilon $-expansions ($%
\epsilon =4-d$) are available.\cite{Duplantier}
\begin{eqnarray}
\gamma _{p}-1=\left( 4+9p(3-p)\right) /64\quad  &&d=2\text{ (exact)}  \notag
\\
\gamma _{p}-1=0\quad  &&d>4\text{ (exact)}  \notag \\
\gamma _{p}-1={\frac{\epsilon }{16}}p(3-p)+o(\epsilon ^{2})\quad
&&d=4-\epsilon   \label{bulkgamma}
\end{eqnarray}%
These estimates allow for an exact determination of $p^{\star }$ in two
dimensions, we get $p^{\star }=5$, the first order $\epsilon $-expansion
happens to give the same value. Assuming that the first order $\epsilon $%
-expansions also gives a fair estimate of $p^{\star }$ in three dimensions
we obtain $p^{\star }=9$. The intermediate regime where only part of the
chains are dissymmetric hence extends over aggregation numbers $5$ to $8$.
Like the flat brush limit, the Daoud and Cotton limit ($p\gg 1$, $\gamma
_{p}\propto -p^{d/(d-1})$) is dominated by dissymmetric configurations (Figure %
\ref{Narms}).

Similar arguments can be developed for sliding surface aggregates with a
small core grafted on an impenetrable wall. The critical exponents $\gamma
_{p}$ have to be replaced by the corresponding surface exponents $\gamma
_{p}^{s}$ in eqs. (\ref{evstar}-\ref{intermaggr}). The following estimates
of the surface exponents can be used:\cite{Duplantier}
\begin{eqnarray}
\gamma _{p}^{s}-1=p(15-18p)/64\quad &&d=2\text{ (exact)}  \notag \\
\gamma _{p}^{s}-1=-p/2\quad &&d>4\text{ (exact)}  \notag \\
\gamma _{p}^{s}-1=-p/2+{\frac{\epsilon }{16}}p(3-p)+o(\epsilon ^{2})\quad
&&d=4-\epsilon  \label{surfacegamma}
\end{eqnarray}
The first order $\epsilon$-expansion gives the estimate $p^{\star s}=3$ in
two dimensions and $p^{\star s}=5$ in three dimensions. The former is to be
compared with the exact value $p^{\star s}=2$ in two dimensions. The first
order epsilon expansion is likely to slightly overestimate $p^{\star s}$
also in three dimensions. If one would use the Daoud and Cotton\cite{Daoud}
like approximation $\gamma _{p}^{s}/\gamma _{2p}=1/2$ which is exact for
infinite $p$, one would get $p^{\star s}=p^{\star }/2$. As expected,
excluded volume correlations are stronger at the surface and favor
dissymmetric configurations. In three dimensions, the intermediate regime is
narrow and covers aggregation numbers $3$ and $4$ on the basis of the above
estimates.

This discussion embodies the special case $p=1$ of a single chain described
in the ideal case earlier. Let again $n$ be the length of the shorter
branch, following eq.(\ref{evstar}), with surface exponents, we can write
the partition function
\begin{equation}
Z(n)=(N-n)^{\gamma _{1}^{s}-1}n^{\gamma _{1}-\gamma _{1}^{s}-1}\quad n<N
\label{evsingle}
\end{equation}
where the exact relation $\gamma _{2}^{s}=\gamma _{1}-1$ between critical
exponents has been used. Eq.(\ref{idealsingle}) is recovered, up to the
unimportant normalization factor if ideal exponents are inserted. At first
order in $\epsilon $, $\gamma _{1}-\gamma _{1}^{s}=1/2+o(\epsilon ^{2})$, a
quadratic interpolation between $d=2$ and $d=4$ gives the estimate, $\gamma
_{1}-\gamma _{1}^{s}=1/2-0.027$, the weak divergence of $Z(n)$ at $n=0$ is
only slightly stronger than in the ideal case.

Sliding grafts on small colloids or sliding star-like micelles illustrate
thus the interesting adaptability of sliding grafted chains. If there are
only a few chains \emph{per} colloid or micelle, they adopt symmetric
configurations and hence all branches participate in the corona. If there
are many chains \emph{per} colloid or micelle, they adopt highly
dissymmetric configurations and only half of the branches participate in the
corona. Interestingly, there is an intermediate regime where a fixed number $%
p^{\star }$ of branches participates in the corona. This suggest that in
this regime fluctuations in the number of grafts, or aggregation number,
could be somehow washed out. Following our estimate based on critical
exponents aggregates comprising from $5$ to $8$ chains would present $9$
longer branches participating in the corona.

\section{Conclusion\label{secconclusion}}

Topological grafts as the ones provided by grafted cyclodextrin-PEO
complexes allow for a new class of materials, where the connection between
the different elements composing the material is defined by simple
topological rules.

One of the most important new features of the SGP layers that we considered
here is that the sliding chains can adapt to external conditions. In the
mushroom regime, where chains are only sparsely grafted to the surface, the
two arms adopt mainly symmetric conformations. We exactly showed this for
ideal chains and checked that it remains true for chains with excluded
volume statistics. In the latter case excluded volume correlations only
slightly increase the probability of dissymmetric configurations. In these
SGP systems, external forces selectively applied to one end, can easily
favor dissymmetric configurations. In densely grafted layers, in contrast,
the chains adopt very asymmetric configurations to accommodate the strong
inter-chain excluded volume interactions. This is merely because the free
energy density of a layer of equal chains increases linearly with chain
length but super linearly with grafting density. Qualitatively, a typical
graft comprises a long branch and a short one filling one correlation volume
(\emph{blob}) at the surface. As the density decreases and the mushroom
regime is approached the two branches become typically comparable in size,
and the size distribution is not longer bimodal. We showed also that a
comparable behavior can be obtained for SGP layers grafted onto curved
surfaces, with perhaps more adaptability due to the extra available space
around the surface. In particular, this leads to an intermediate grafting
density regime where coexist symmetric and asymmetric chain configurations.

We believe that SGP layers represent a completely new type of interfacial
polymer structures and as such, open many new possibilities that we have
barely considered here. For instance, we expect the steric forces between
SGP layers to be qualitatively different from the usual steric repulsion
between grafted polymer layers. We hope to address this and other related
questions in future extensions of our work.

\acknowledgements{ V. B. gratefully acknowledges the French Space
Agency, CNES for a research post-doctoral fellowship.}


\begin{thebibliography}{99}
\bibitem{Nakashima} Nakashima, N.; Kawabuchi, A.; Murakami, H. \emph{J. Inc.
Phen. Mol. Rec. Chem.} \textbf{1998}, \emph{32}, 363--373.

\bibitem{Ogino} Ogino, H. \emph{J. Am. Chem. Soc.} \textbf{1981}, \emph{103}%
, 1303.

\bibitem{OginoOhata} Ogino, H.; Ohata, K. \emph{Inorg. Chem.} \textbf{1984},
\emph{23}, 3312.

\bibitem{PGGSliding} de~Gennes, P.-G. \emph{Physica A} \textbf{1999}, \emph{%
271}, 231--237.

\bibitem{Cacialli} Cacialli, F.; Wilson, J.~S.; Michels, J.~J.; Daniel, C.;
Silva, C.; Friend, R.~H.; Severin, N.; Samor{\`i}, P.; Rabe, J.~P.;
O'connell, M.~J.; Taylor, P.~N.; Anderson, H.~L. \emph{Nature Materials}
\textbf{2002}, \emph{1}, 160.

\bibitem{Ballardini} Ballardini, R.; Balzani, V.; Credi, A.; Gandolfi,
M.~T.; Venturi, M. \emph{Acc. Chem. Res.} \textbf{2001}, \emph{34}, 445--455.

\bibitem{HaradaAcc} Harada, A. \emph{Acc. Chem. Res.} \textbf{2001}, \emph{34%
}, 456--464.

\bibitem{Schalley} Schalley, C.~A.; Beizai, K.; Gtle, F.~V. \emph{Acc. Chem.
Res.} \textbf{2001}, \emph{\ 34}, 465--476.

\bibitem{Tamura} Tamura, M.; Gao, D.; Ueno, A. \emph{Chem. Eur. J.} \textbf{%
2001}, \emph{7}(7), 1390--1397.

\bibitem{HaradaCoo} Harada, A. \emph{Coord. Chem. Rev.} \textbf{1996}, \emph{%
148}, 115--133.

\bibitem{Wei} Wei, M.; Shin, I.~D.; Urban, B.; Tonelli, A.~E. \emph{J. Pol.
Sci. Part B: Pol. Phys.} \textbf{2004}, \emph{42}, 1369--1378.

\bibitem{Rekharsky} Rekharsky, M.~V.; Inoue, Y. \emph{Chem. Rev.} \textbf{%
1998}, \emph{98}, 1875--1917.

\bibitem{Shridhar} Heddarimane, S.~M.; Fleury, G.; Schlatter, G.; Brochon,
C.; Hadziioannou, G.; Marques, C.~M. preprint.

\bibitem{CohenSt} Fleer, G.~J.; Stuart, M.~C.; Scheutjens, J.; Cosgrove, T.;
Vincent, B. \emph{\ Polymer at Interfaces;} \newblock Chapman et Hall:
London, 1993.

\bibitem{Lipowsky} Lipowsky, R.; Sackmann, E. \emph{Structure and Dynamics
of Membranes;} \newblock Elsevier, 1995.

\bibitem{SA} Alexander, S. \emph{Journal de Physique (France)} \textbf{1977}%
, \emph{38}, 983--987.

\bibitem{PGG} de~Gennes, P.-G. \emph{Scaling Concepts in Polymer Physics;} %
\newblock Cornell University Press: Ithaca and London, second ed., 1985.

\bibitem{HTL} Halperin, A.; Tirrell, M.; Lodge, T.~P. \emph{Adv. Polym. Sci.}
\textbf{1992}, \emph{\ 100}, 31.

\bibitem{Milner} Milner, S.~T.; Witten, T.~A.; Cates, M.~E. \emph{%
Macromolecules} \textbf{1989}, \emph{22}, 853--861.

\bibitem{Auvray} Auroy, P.; Auvray, L.; Leger, L. \emph{Phys. Rev. Lett.}
\textbf{1991}, \emph{66}, 719.

\bibitem{Doi} Doi, M.; Edwards, S.~F. \emph{The theory of polymer dynamics;} %
\newblock Clarendon Press: Oxford, 1986.

\bibitem{foot1} The image of eq. (\ref{Zn1}) in the Laplace space yields $
\tilde{Z}(s)\sim \frac{1}{s}\prod_{k=2}^{m}\frac{1}{\beta _{k-1}^{3}}\exp %
\left[ -\beta _{k-1}\sqrt{s}\right]\left( 1+\beta _{k-1}\sqrt{s}\right) $.
Applying the equality $ \lim\limits_{N\rightarrow \infty
}Z=\lim\limits_{s\rightarrow 0}s\tilde{Z}(s) $ we obtain eq. (\ref{Z}).

\bibitem{Abramowitz} Abramowitz, M.; Stegun, I.~A. \emph{Handbook of
mathematical functions;} \newblock National Bureau of Standards: Washington
DC, 1964.

\bibitem{foot2} The image of eq. (\ref{UP}) in the Laplace space is $\tilde{P%
}(s)=-1/(swS_{0})$. Combining this expression with the image of eq. (\ref{Sn}%
) $\tilde{S}(s)=-S_{0}\tilde{P}(s)/s$ and applying the inverse Laplace
transform we get eq.(\ref{Un}).

\bibitem{Duplantier} Duplantier, B. \emph{J. Stat. Phys.} \textbf{1989},
\emph{54}, 581--680.

\bibitem{Diez} Johner, A.; Joanny, J.-F.; Orrite, S.~D.; Avalos, J.~B. \emph{%
Europhys. Lett.} \textbf{2001}, \emph{56}, 549.

\bibitem{Daoud} Daoud, M.; Cotton, J.~P. \emph{J. Phys. (France)} \textbf{%
1982}, \emph{43}, 531.
\end{thebibliography}

\end{document}